\documentclass[aps,prb,twocolumn,groupeauthor,showpacs]{revtex4}

\usepackage{amsmath}
\usepackage[pdftex]{graphicx}
\usepackage{dcolumn}
\usepackage{bm}
\usepackage{ulem}
\usepackage{multirow}

\begin{document}

\title{Ultrafast laser-induced changes of the magnetic anisotropy\\in a low-symmetry iron garnet film}

\author{L.\,A.\,Shelukhin, V.\,V.\,Pavlov, P.\,A.\,Usachev, P.\,Yu.\,Shamray, R.\,V.\,Pisarev, A.\,M.\,Kalashnikova \email{kalashnikova@mail.ioffe.ru}}
\affiliation{Ioffe Institute, 194021 St. Petersburg, Russia}

\date{\today}

\begin{abstract}
We explore a thermal mechanism of changing the anisotropy by femtosecond laser pulses in dielectric ferrimagnetic garnets by taking a low symmetry (YBiPrLu)$_3$(FeGa)$_5$O$_{12}$ film grown on the (210)-oriented Gd$_3$Ga$_5$O$_{12}$ substrate as a model media. We demonstrate by means of spectral magneto-optical pump-probe technique and phenomenological analysis, that the magnetization precession in such a film is triggered by laser-induced changes of the growth-induced magnetic anisotropy along with the well-known ultrafast inverse Faraday effect. The change of magnetic anisotropy is mediated by the lattice heating induced by laser pulses of arbitrary polarization on a picosecond time scale. We show that the orientation of the external magnetic field with respect to the magnetization easy plane noticeably affects the precession excited via the anisotropy change. Importantly, the relative contributions from the ultrafast inverse Faraday effect and the change of different growth-induced anisotropy parameters can be controlled by varying the applied magnetic field strength and direction. As a result, the amplitude and the initial phase of the excited magnetization precession can be gradually tuned.
\end{abstract}


\maketitle

\section{Introduction}

Ferrimagnetic rare-earth iron garnets $R_3$Fe$_5$O$_{12}$ and related compounds, where $R$ stands for yttrium, rare-earth and some other ions, e.g. bismuth, are highly resistive dielectrics with the bandgap of $E_g\sim$2.8\,eV.
These materials have passed through several periods of the strong research interest, triggered by their unique physical properties and important applications.\cite{Winkler-book,Paoletti-book}
First of all, yttrium iron garnet (YIG) possesses a record-narrow width of the ferromagnetic resonance line\cite{Gurevich-book} and the strong magneto-acoustic coupling,\cite{Spencer} thus being a basic medium for a large family of microwave devices.
Being centrosymmetric cubic in the bulk, epitaxial magnetic garnet films reduce their crystallographic symmetry and generally lose the center of inversion, which allows for various effects forbidden in cubic bulk samples, such as the optical second harmonic generation,\cite{Pavlov-PRB1997,Gridnev-PRB2001} and intrinsic giant linear magneto-electric effect.\cite{Krichevtsov-JETPLett1989,Logginov-JETPLett2007,Popov-PRB2014}
Owing to the piezomagnetic response, thin magnetic garnet films are among the building blocs for composite multiferroics as well.\cite{Fetisov-APL2006}
High values of Faraday rotation in Bi-substituted YIG allow designing efficient magneto-optical isolators and waveguides.\cite{Tien-APL1972}
The same property makes magnetic garnets the key material for engineering magneto-photonic\cite{Inoue-JPD2006} and magneto-plasmonic\cite{Armelles-AOM2013} structures.
Thin magnetic garnet films with uniaxial anisotropy were among the model media for developing magnetic bubble domain technology,\cite{Eschenfelder-book,Malozemoff-Book1979} and became of interest recently again owing to certain analogies between bubble domains and skyrmions,\cite{Ogawa}, and to possibility of controlling domain walls by localized electric fields.\cite{Pyatakov-UFN2015, Khokhlov-SciRep2017}
Nowadays, thin YIG films are the model functional media for testing various concepts of magnonics,\cite{Kruglyuk-JPD2010} owing to their exceptionally low spin waves damping,\cite{Serga-JPD2010} and novel effects at the dielectric garnet/metal interfaces.\cite{Nakayama-PRL2013}

This outstanding functionality of the magnetic garnet films originates from the fact, that their magnetization, magnetic anisotropy, compensation points and other properties can be tailored in a wide range.
Thus, growth conditions, a type of substrate and a chemical composition allows fabricating the garnets with easy-plane, out-of-plane, and more intricate types of the anisotropy.\cite{Winkler-book,Paoletti-book}
Furthermore, the magnetic anisotropy of garnets is highly susceptible to various external stimuli, such as temperature,\cite{Tekielak-JPhys1997} strain,\cite{Fetisov-APL2006} and optical irradiation.\cite{Fedorov-JETP1989,Chizhik-PRB1998}
As a result, efficient dynamical modulation of parameters of ferromagnetic resonance and spin waves spectra, domain patterns, etc. can be realized by using these external stimuli.
Recently, the control of the magnetic anisotropy of garnet films was demonstrated by means of femtosecond laser pulses, thus showing the feasibility of ultrafast photomagnetic effects.\cite{Hansteen-PRL2005}

By now, femtosecond laser-induced changes of the magnetic anisotropy has been shown to be one of the most common effects, observed in magnetically-ordered dielectrics,\cite{Hansteen-PRL2005,Kimel-Nature2004,Caretta-PRB2015,Stupakiewicz-arxiv2016} semiconductors\cite{Hashimoto-PRL2008,Scherbakov-PRL2010} and metals.\cite{Koopmans-PRL2002,Carpene-PRB2010,Kats-PRB2016}
In all substances ultrafast change of the intrinsic magnetocrystalline and shape anisotropies by femtosecond laser pulses or appearance of a transient anisotropy axis manifest themselves via coherent spin precession, despite very different microscopic mechanisms underlying these processes, related, first of all, to the corresponding electronic band structures.
For example, excitation by a femtosecond laser pulse of magnetically ordered metals results in subpicosecond increase of the electronic temperature which determines subsequent dynamics of other subsystem.\cite{Wilks-JAP2005,Hohlfeld-PRL1997,Stamm-NatureMat2007}
Therefore, in such media changes of the shape and magnetocrystalline anisotropies mostly result from the ultrafast heating\cite{Koopmans-PRL2002,Carpene-PRB2010} and related effects.\cite{Kats-PRB2016}
By contrast, magnetically ordered dielectrics subjected to femtosecond laser pulses demonstrate a variety of both thermal and nonthermal mechanisms leading to the anisotropy changes.

The most prominent nonthermal changes of the magnetic anisotropy in dielectrics under the action of laser pulses were observed in several substituted iron garnets,\cite{Hansteen-PRL2005,Atoneche-PRB2010,Yoshimine-JAP2014,Stupakiewicz-arxiv2016} in which linearly polarized pulses induce a transient anisotropy axis due to charge-transfer optical transitions.
These results demonstrated feasibility of controlling the magnetic anisotropy by changing the azimuthal angle of the laser pulse polarization.
Furthermore, this ultrafast photomagnetic effect allowed achieving relatively high amplitudes of laser-induced magnetization precession\cite{Atoneche-PRB2010} and even controllable switching of magnetization.\cite{Stupakiewicz-arxiv2016}
Recently, impulsive photomagnetic effect has been discussed in Ref.\,\onlinecite{Koene-PRB2015} as an another candidate for triggering the magnetization precession in a magnetic garnet, which microscopical picture is yet to be understood. The laser-induced uniaxial anisotropy mediated by the acoustic phonons has been shown to enhance magnetization of the Cu-based organic-inorganic Heisenberg magnets.\cite{Caretta-PRB2015}

Rapid heating related to the laser pulses can also lead to a modification of the intrinsic magnetocrystalline anisotropy of dielectrics.
Exploring this mechanism in the rare-earth orthoferrites in the vicinity of spin reorientation phase transitions\cite{Kimel-Nature2004} yielded a number of remarkable results on coherent control of magnetization.\cite{Kimel-NaturePhys2009,deJong-PRL2012,Afanasiev-PRL2016}
Evidently, laser-induced thermal changes of the magnetic anisotropy should be a general phenomenon in dielectrics, not restricted exclusively to the vicinity of phase transitions.
Such a process, therefore, can be an alternative way to the spin waves excitation.
Thermal modulation of the anisotropy can affect the spin waves spectrum and other dynamical properties and, therefore, understanding timescales and strength of this effects in dielectrics is important for implementation of laser pulses as excitation stimuli in future magnonic,\cite{Satoh-NatureP2012} magneto-plasmonic,\cite{Belotelov-PRB2012,Bossini-ACSPoton2016} spintronic and spin-optronic devices.
However, to the best of our knowledge ultrafast thermal changes of magnetic anisotropy of dielectrics, except for orthoferrites,\cite{Kimel-Nature2004,Kimel-NaturePhys2009,deJong-PRL2012,deJong-PRB2011,Kimel-PRB2006,Afanasiev-PRL2016} have not been explored so far.

In this paper we report on the results of the experimental studies of ultrafast magnetization dynamics in a low-symmetry substituted iron garnet film characterized by a pronounced growth-induced anisotropy of the easy-axis type.
We demonstrate that the impact of a femtosecond laser pulse on the garnet film induces polarization-independent changes of its growth-induced anisotropy parameters. Their relative contributions can be distinguished by analysing the azimuthal field dependencies of the initial phase of the induced precession owing to the low symmetry. We argue that the most plausible mechanism underlying the observed change of the magnetic anisotropy is the lattice heating, which is expected to take place on a picosecond timescale. We show that the amplitude of the magnetization precession excited via this mechanism is comparable to that induced by the ultrafast inverse Faraday effect. We show that the relative contribution of these two mechanisms of the magnetization precession excitation can be tuned by changing the value of the applied magnetic field. Importantly, this allowed us to vary gradually the initial phase of the magnetization precession in the range of $\sim\pi/2$.

This paper is organized as follows. In Sec.\,\ref{Sec-anisotropy} we introduce phenomenological description of the magnetic anisotropy of the substituted iron garnet films grown on (210) substrate. We consider how the ultrafast change of the magnetic anisotropy parameters is expected to affect the magnetic state of the low symmetry magnetic garnet film. Sec.\,\ref{Sec-samples} is dedicated to the characterization of the magnetic garnet film chosen for the study.
In Sec.\,\ref{Sec-experimental} we present the details of the magneto-optical pump-probe experiments. In Sec.\,\ref{Sec-precession} the experimental data on the magnetization dynamics after the laser pulse excitation in the (210) garnet film are discussed. This is followed by the analysis of two mechanisms of the precession excitation, the ultrafast inverse Faraday effect (Sec. \ref{Sec-IFE}) and change of the growth-induced anisotropy parameters under the influence of laser pulses (Sec. \ref{Sec-inducedAnisotropy}).

\section{Magnetic anisotropy of a garnet film grown on a (210)-oriented substrate}\label{Sec-anisotropy}

Substituted iron garnet films are characterized by the magnetic anisotropy which is the consequence of the interplay between cubic anisotropy inherent to their crystallographic structure and the growth- and stress-induced ones.
The latter are dependent on several factors such as substrate lattice parameters and crystallographic orientation, particular chemical composition and parameters of the growth technology (for a review see e.g. Ref.\onlinecite{Eschenfelder-book}).
Typically in films grown on low symmetry substrates the growth- and/or stress-induced contributions dominate, which results in magnetic anisotropy of the easy-axis type.

\begin{figure}
\includegraphics[width=6cm]{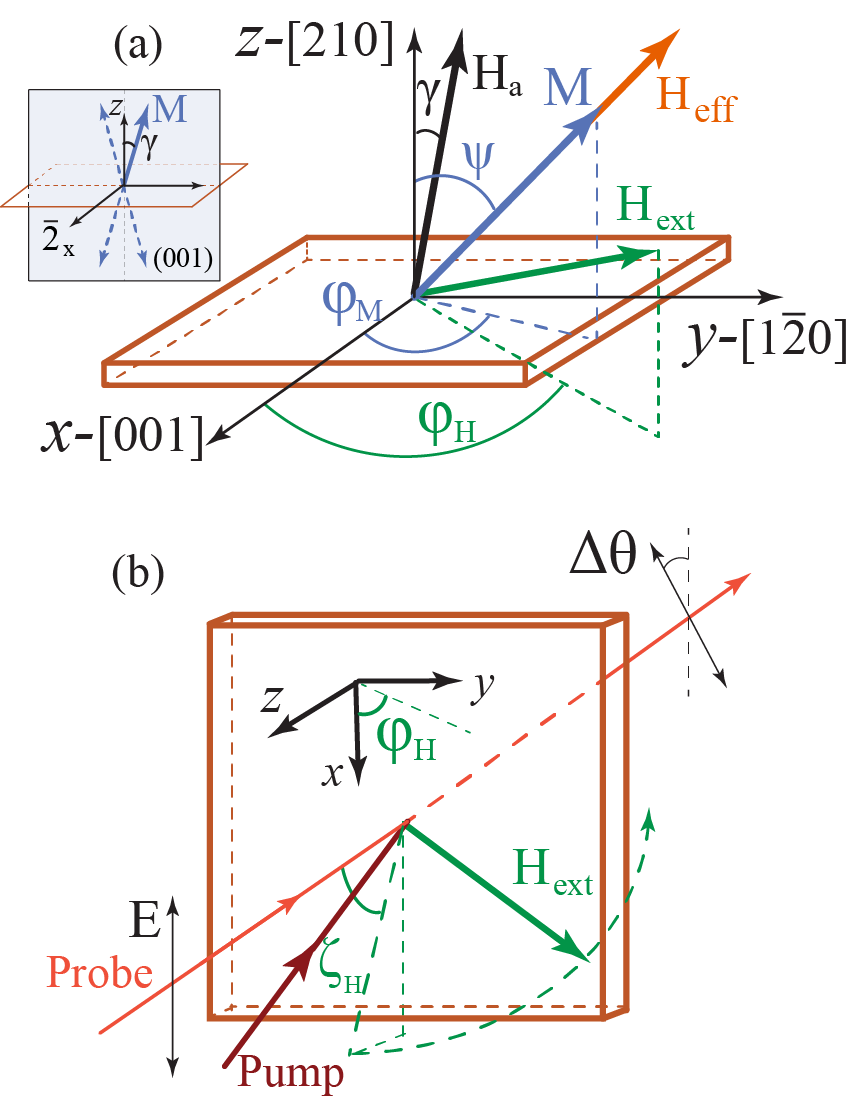}
\caption{(Color online) (a) Orientations of the crystallographic axes of the (210)-film, the applied magnetic field $\mathbf{H}_\mathrm{ext}$, the effective anisotropy field $\mathbf{H}_\mathrm{a}$, and the net effective field $\mathbf{H}_\mathrm{eff}$. $x,y,z$ axes are directed along [001], [1$\bar{2}0$], and [210] crystallographic axes, respectively. Effective anisotropy field $\mathbf{H}_\mathrm{a}$ makes an angle $\gamma=16^\mathrm{o}$ with the $z$-axis in the ($yz$)-plane, as shown in the inset.\cite{Krichevtsov-JETPLett1989} The magnetic field direction is described by the polar angle $\psi_\mathrm{H} = 80^\mathrm{o}$ and the azimuthal angle $\varphi_\mathrm{H}$. The direction of magnetization $\mathbf{M}$ is described by the polar $\psi_M$ and azimuthal $\varphi_\mathrm{M}$ angles. In a general case the magnetization $\mathbf{M}$ is not collinear with the applied magnetic field and is not in the ($yz$)-plane. (b) Geometry of the pump-probe experiment (see text for the details).}
\label{Fig:Geometry}
\end{figure}

We consider here a magnetic garnet film grown on the (210)-substrate.
The reference frame is chosen as shown in Fig.\ref{Fig:Geometry}(a).
The $x$-axis is directed along the [001] crystallographic direction.
From the symmetry point of view this direction is the $\bar{2}_x$ axis and the crystallographic point group of this film is $m$.\cite{Gridnev-PRB2001}
The $z$-axis is directed along the [210] crystallographic axis and is normal to the sample plane.
The magnetic anisotropy energy of such a film can be expressed as\cite{Eschenfelder-book,Nistor-JAP2007} (see also Appendix\,\ref{App-anisotropy})
\begin{equation}
w_\mathrm{a}=K_um^2_z+K_im^2_y+K_{yz}m_ym_z+w_\mathrm{cub},\label{Eq:anisotropy}
\end{equation}
where $m_{k}=M_k/M_S$ $(k=x,y,z)$ are the normalized components of the magnetization $\mathbf{M}$, $M_S$ is the saturation magnetization.
In this expression $w_\mathrm{cub}$ is the cubic anisotropy energy, $K_u$ and $K_i$, and $K_{yz}$ are the uniaxial out-of-plane and in-plane, and orthorhombic anisotropy parameters, respectively. The first three terms in Eq.\,(\ref{Eq:anisotropy}) have two contributions, the growth-induced and the stress-induced ones. Relative strength of these contributions varies depending on a composition of a particular films, film/substrate lattice mismatch, and the growth conditions. In particular, in Bi-substituted iron garnets, which are investigated in our experiments, the growth-induced anisotropy due to Bi$^{3+}$ ions occupying dodecahedral sites dominates over the stress-induced one,\cite{Hansen-JAP1985} as we also discuss in more details in the Appendix\,\ref{App-anisotropy}. The cubic anisotropy $w_\mathrm{cub}$ is typically much weaker than the growth-induced one, and is omitted in the following consideration.

The equilibrium orientation of the magnetization is determined by the ratios between growth-induced anisotropy parameters entering Eq.\,(\ref{Eq:anisotropy}).
In particular, the $yz$ plane is the easy plane of the magnetization, given $K_{yz}<0$.
The orientation of the magnetization in this plane is determined by the values of $K_u$, $K_i$ and $K_{yz}$ for a particular film.
The expression for the effective growth-induced anisotropy field $\mathbf{H}_\mathrm{a}$ has a form
\begin{equation}
\mathbf{H}_\mathrm{a}=-\frac{\partial w_\mathrm{a}}{\partial\mathbf{M}}=\frac{1}{M^2_S}\left(\begin{array}{c}
                           0\\
                           -2K_iM_y-K_{yz}M_z\\
                           -2K_uM_z-K_{yz}M_y
                         \end{array}\right).\label{Eq:Heff}
\end{equation}
From this expression one can see that a laser-induced change of the anisotropy parameters $K_u$\, $K_i$ or $K_{yz}$ should lead to changes of both the value and the direction of the effective anisotropy field.
In order to illustrate how changes of these parameters can lead to the excitation of the magnetization precession we employ the phenomenological approach based on the Landau-Lifshitz equation\cite{Landau-Lifshitz-1935}
\begin{equation}
\frac{d\mathbf{M}}{dt}=-\gamma\mathbf{M}\times\mathbf{H}_\mathrm{eff}=-\gamma\mathbf{M}\times(\mathbf{H}_\mathrm{ext}+\mathbf{H}_\mathrm{a}+\mathbf{H}_\mathrm{d}),\label{Eq:LL}
\end{equation}
where $\gamma$ is the gyromagnetic ratio, $\mathbf{M}$ is the magnetization, and $\mathbf{H}_\mathrm{eff}$ is the net effective magnetic field, which consists of the applied magnetic field $\mathbf{H}_\mathrm{ext}$, the internal anisotropy field $\mathbf{H}_\mathrm{a}$, and the demagnetizing field $\mathbf{H}_\mathrm{d}=-4\pi M_z\mathbf{z}$.
In Eq.\,(\ref{Eq:LL}) the term responsible for the precession damping is omitted, since we focus on a time scale when its contribution can be ignored.

In this approach the effect of the femtosecond laser pulse is introduced into the expression for the effective magnetic field $\mathbf{H}_\mathrm{eff}$ by adding new terms and modifying existing ones.
In particular, (i) the change of the anisotropy constants $\Delta K_u(t)$, $\Delta K_i(t)$, and $\Delta K_{yz}(t)$ manifests itself in the change of the effective anisotropy field $\Delta\mathbf{H}_\mathrm{a}(t)$; (ii) the effective field $\mathbf{H}_\mathrm{om}(t)$ accounts for the opto-magnetic effects, such as the ultrafast inverse Faraday and Cotton-Mouton effects.\cite{Kimel-Nature2005,Kalashnikova-PRL2007,Kalashnikova-PRB2008,Gridnev-PRB2008}

Using Eqs.\,(\ref{Eq:Heff}, \ref{Eq:LL}) one can derive general expressions for the torque acting on the magnetization at $t=0$ due to these effects, assuming that the anisotropy is modified at the sufficiently short time scale:
\begin{widetext}
\begin{eqnarray}
\left.\frac{1}{\gamma}\frac{d\mathbf{M}}{dt}\right|_{t=0}&=&-\mathbf{M}\times(\Delta\mathbf{H}_\mathrm{a}+\mathbf{H}_\mathrm{om})\label{Eq:Torque}\\
&=&                          \left(\begin{array}{c}
                           (2(\Delta K_i-\Delta K_u)m_ym_z+\Delta K_{yz}(m^2_z-m^2_y))\\
                           (2\Delta K_um_xm_z+\Delta K_{yz}m_xm_y)\\
                           (-2\Delta K_im_xm_y-\Delta K_{yz}m_xm_z)
                         \end{array}\right)+
                         \left(\begin{array}{c}
                           (H_{\mathrm{om}z}M_y-H_{\mathrm{om}y}M_z)\\
                           (H_{\mathrm{om}x}M_z-H_{\mathrm{om}z}M_x)\\
                           (H_{\mathrm{om}y}M_x-H_{\mathrm{om}x}M_y)
                         \end{array}\right)\nonumber
\end{eqnarray}
\end{widetext}
Importantly, Eq.\,(\ref{Eq:Torque}) clearly demonstrates that the strength and the direction of the torque acting on the magnetization due to the induced changes of any anisotropy constants $\Delta K$  as well as due to ultrafast opto-magnetic effects are determined by the initial orientation of the magnetization.

Dependence of the torque (\ref{Eq:Torque}) on the initial magnetization direction makes iron garnet films grown on (210)-oriented substrates the model media to explore experimentally laser-induced changes of the magnetic anisotropy parameters $\Delta K$. Indeed, the external magnetic field $\mathbf{H}_\mathrm{ext}$ of tunable strength applied along the hard axis $x$ gradually changes the equilibrium orientation of the magnetization.
Therefore, this should favor experimental detection of the magnetization dynamics excited due to the changes of the anisotropy parameters and, moreover, can allow one to determine their relative contributions.
It is also important to emphasize that the character of the magnetization dynamics excited due to the such changes would be strongly dependent on the time scale of the involved processes.
Generally speaking, it should occur on a time scale shorter than the characteristic time of the magnetization precession.
If it is not the case, then one can expect slow deflection movement of the magnetization towards a new equilibrium position instead of the precessional motion.

We note that, when deriving the expression for the torque (\ref{Eq:Torque}), we neglected a possible change of the saturation magnetization $\Delta M_S(t)$ due to the laser-induced demagnetization.
This is justified because the optically-driven demagnetization in dielectrics triggered by optical pulses occurs on a time scale of several hundreds of picoseconds,\cite{Kimel-PRL2002,Ogasawara-PRL2005} which is longer than the typical period of ~100\,ps of precession in magnetic garnet films.
Therefore, the demagnetization, although resulting in the changes of demagnetizing field $\Delta\mathbf{H}_\mathrm{d}$, is not expected to contribute to the torque (\ref{Eq:Torque}). The validity of this assumption was verified experimentally as discussed in the Appendix\,\ref{App-demagnetization}.

\section{Samples}\label{Sec-samples}

For investigating the feasibility of the described above laser-induced modification of anisotropy parameters we have chosen Bi-substituted iron garnet film (Y$_{0.99}$Bi$_{1.64}$Pr$_{0.25}$Lu$_{0.23}$)(Fe$_{3.75}$Ga$_{1.16}$)O$_{12}$ grown by liquid phase epitaxy method on (210)-oriented substrate of a gadolinium gallium garnet Gd$_3$Ga$_5$O$_{12}$(GGG). The film composition was verified by the X-ray fluorescence measurements. The film thickness was of 10\,$\mu$m. X-ray diffraction characterization yielded the lattice constants of $a^f=$12.5322\,\AA \, and $a^s$=12.4844\,\AA \, for the garnet film and the GGG substrate, respectively. The lattice mismatch between the film and the substrate, introduced according to Ref.\,\onlinecite{Eschenfelder-book}, is of $\Delta a/a=(a^s-a^f)/a^f=$-0.38\,\%. Effective anisotropy easy axis is oblique in the $yz$ plane with the angle of 16$^\mathrm{o}$ to the sample normal,\cite{Krichevtsov-JETPLett1989} as shown in Fig.\,\ref{Fig:Geometry}(a). Saturation magnetization is $M_s=10^4$\,A/m.\cite{Krichevtsov-JETPLett1989}

\begin{figure}
\includegraphics[width=6cm]{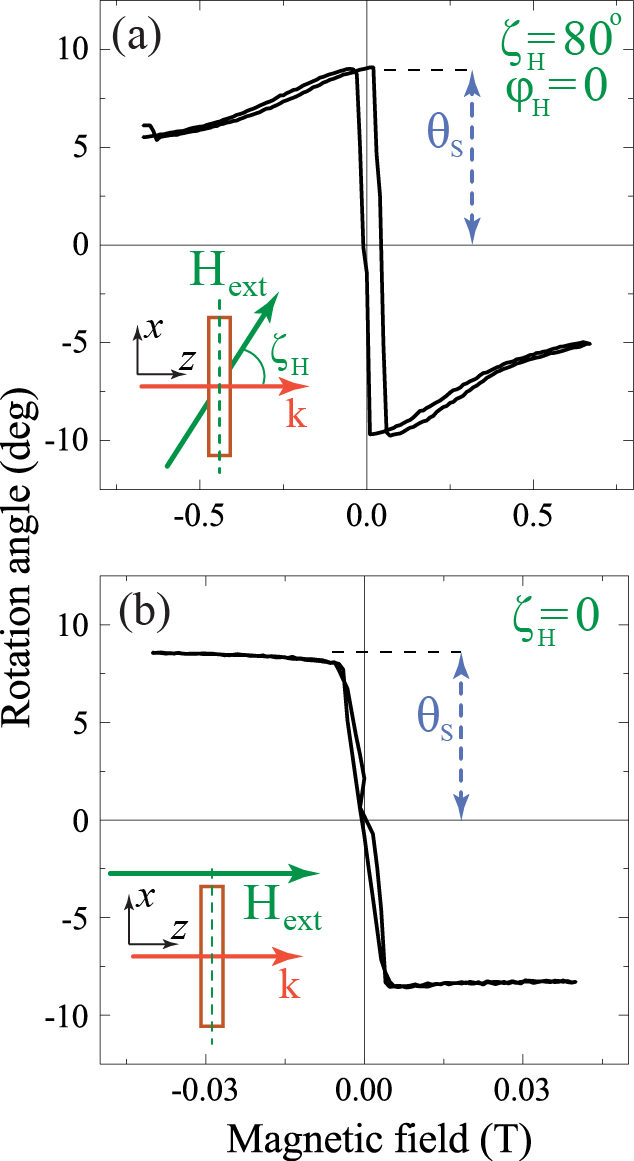}
\caption{(Color online) Static Faraday rotation measured as a function of the external magnetic ﬁeld applied at (a) $\zeta_\mathrm{H} = 80^\mathrm{o}$ and (b) $\zeta_\mathrm{H} = 0$ to the sample normal (as shown in insets). $\theta_s$ denotes the Faraday rotation at remanence, which is proportional to the samples magnetization $M_s$.}
\label{Fig:Faraday}
\end{figure}

The sample was further characterized by means of magneto-optical Faraday rotation measurements at a photon energy of 1.8\,eV. The rotation angle $\theta$ of the polarization plane of the light propagating along the sample normal was measured as a function of the DC magnetic field $H_\mathrm{ext}$. In this geometry the measured Faraday rotation $\theta$ is proportional to the out-of-plane, or $M_z$, component of the film magnetization.
Fig.\,\ref{Fig:Faraday} shows the field dependencies of the Faraday rotation measured when the field was applied at 80$^\mathrm{o}$ with respect to the sample normal (a) and along the latter (b). The Faraday rotation $\theta_s$ at remanence is of 8$^{\circ}$, which is consistent with high level of Bi-substitution resulting in increase of the magnetooptical susceptibility as compared to unsubstituted YIG.\cite{Hansen-PRB1983}
Presented data also confirm that the magnetization easy axis is oriented close to the sample normal. As one can see from Fig.\,\ref{Fig:Faraday}(a), the angle between the magnetization and the easy axis is gradually increasing with increase of the field applied at large angle to the sample normal.

In order to estimate the magnetic anisotropy strength of the studied samples we performed ferromagnetic resonance measurements. However, due to the large FMR linewidth no reliable numbers could be extracted from the measurements. Therefore, we have obtained the magnetic anisotropy parameters of the studied film from the field dependences of the Faraday rotation (Fig.\,\ref{Fig:Faraday}(a)) and of the frequency of the magnetization precession excited in our pump-probe experiments. Magnetic anisotropy parameters were found to be $K_u\sim-5\cdot10^{3}$J/m$^2$; $K_i\sim-3\cdot10^{3}$J/m$^2$; $K_{yz}\sim-8.7\cdot10^{3}$\,J/m$^2$. These parameters have contributions (see Appendix\,\ref{App-anisotropy} for details) from the growth-induced ($K_u^g\approx-7\cdot10^3$\,J/m$^3$, $K_i^g\approx-3.5\cdot10^3$\,J/m$^3$, $K_{yz}^g\approx-7.7\cdot10^3$\,J/m$^3$) and the stress-induced ones ($K_u^s\approx2\cdot10^3$\,J/m$^3$, $K_i^s\approx5\cdot10^2$\,J/m$^3$, $K_{yz}^s\approx-1\cdot10^3$\,J/m$^3$).
The growth-induced contribution to the magnetic anisotropy dominates in agreement with previous studies on Bi:YIG with high levels of Bi substitution.\cite{Hansen-JAP1985}

\section{Experimental}\label{Sec-experimental}

All-optical pump-probe experiments were performed employing technique analogous to that described elsewhere.\cite{Kirilyuk-RMP2010}
Optical parametric amplifier (OPA) pumped by the femtosecond regenerative $\mathrm{Yb:KGd(WO_{4})_{2}}$ amplifier (RA) produced 170\,fs laser pulses with repetition rate of 5\,kHz. Most of the experiments were performed with pump and probe central photon energies of $E_\mathrm{ph}$=1.8\,eV.
For the spectrally-resolved studies the pulses with $E_\mathrm{ph}$=1.2\,eV generated directly by the RA were used as the probe pulses.
In the latter experiment the central photon energy of the pump pulses generated by OPA was tuned between 1.7 and 2.0\,eV.

All the measurements were done in transmission geometry, as shown in Fig.\,\ref{Fig:Geometry}(b).
Pump pulses were either linearly or circularly polarized. The angle of incidence for the pump pulses was of 12$^\mathrm{o}$. Pump spot size at the sample was of $150\,\mu$m and the pump fluence was of 7\,mJ/cm$^2$. The pump-induced magnetization dynamics was monitored by measuring the change of the Faraday rotation for the probe pulses as a function of the pump-probe time delay $t$. Linearly polarized probe pulses were incident along the $z$-axis. Probe pulses were focused at the sample to the spot somewhat smaller than the pump. The probe fluence was $\sim$50 times lower than that of the pump.

In this geometry a rotation of the probe polarization plane $\Delta\theta$ is proportional to the change of the magnetization component $M_z$.
Therefore, $\Delta\theta$ normalized by the static Faraday rotation $\theta_s$ at remanence (Fig.\,\ref{Fig:Faraday}) is the measure of the laser-induced out-of-plane deviation of the magnetization from its equilibrium orientation. The external magnetic field $\mathbf{H}_\mathrm{ext}$ of up to 0.63\,T was applied at $\zeta_H=$80$^\mathrm{o}$ to the sample normal in order to deflect the magnetization from its easy axis. The azimuthal angle of the applied field $\varphi_H$ could be varied between 0 and 180$^\mathrm{o}$. All measurement were performed at $T$=295\,K.

In our experiments the dynamics of the magnetization following the femtosecond laser pulse excitation was studied as a function of the pump polarization and of the sign and magnitude of the applied magnetic field $\pm H_\mathrm{ext}$. In order to distinguish helicity-dependent (hd) effects sensitive to the helicity $\sigma^\mathrm{\pm}$ of the pump pulses and the those dependent on the $H_\mathrm{ext}$ sign (fd) we used the expressions
\begin{eqnarray}
\frac{\Delta\theta_\mathrm{hd}}{\theta_s}&=&\frac{\Delta\theta(\sigma^\mathrm{+};+H)-\Delta\theta(\sigma^\mathrm{-};+H)}{2\theta_s};\label{Eq:HD}\\
\frac{\Delta\theta_\mathrm{fd}}{\theta_s}&=&\frac{\Delta\theta(p;+H)-\Delta\theta(p;-H)}{2\theta_s}\label{Eq:FD},
\end{eqnarray}
where $p$ stands for particular pump polarization, and spans from -1 to 0 and to +1 for the left-handed, linearly- and right-handed polarized pump pulses.

\section{Results and discussion}\label{Sec-results}

\subsection{Laser-induced magnetization precession}\label{Sec-precession}

\begin{figure}
\includegraphics[width=8.6cm]{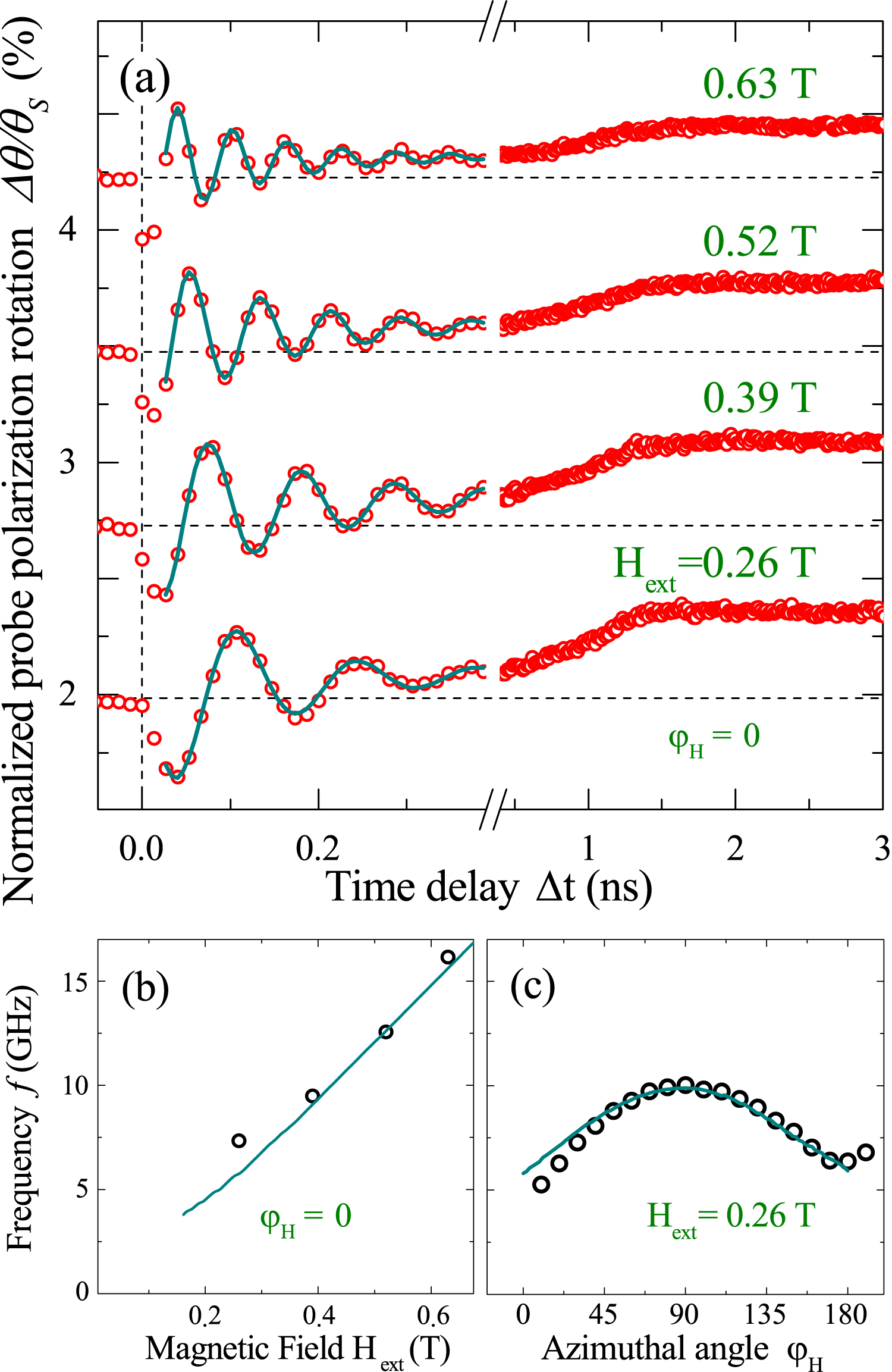}
\caption{(a) Normalized rotation of the probe polarization induced by the left-handed ($\sigma^\mathrm{-}$) circularly polarized pump pulses in the MA-film, measured as a function of the time delay $t$ for various magnitudes of the applied magnetic field. (b) Frequency of oscillations shown in the panel (a) as a function of the magnetic field magnitude (azimuthal angle $\varphi_\mathrm{H}$=0). (c) Frequency of oscillations shown in the panel (a) as a function of the azimuthal angle $\varphi_\mathrm{H}$.}
\label{Fig:Precession}
\end{figure}

Fig.\,\ref{Fig:Precession}(a) shows the rotation of the probe polarization induced by the circularly polarized laser pulse as a function of the pump-probe time delay for different values of the magnetic field $H_\mathrm{ext}$.
Clear oscillations of the probe polarization are observed superimposed on a slowly changing background.
The frequency\,$f$ of the oscillations increases with the field (Fig.\,\ref{Fig:Precession}(b)).
This indicates that the observed oscillations result from the precession of the magnetization triggered by the pump pulses.
Fig.\,\ref{Fig:Precession}(c) shows precession frequency\,$f$ as a function of the azimuthal angle $\varphi_{H}$ of the external field $H_\mathrm{ext}$=0.26\,T (see Fig.\,\ref{Fig:Geometry}(a)).
From these results we were able to determine the orientation of the hard magnetization $x$ axis of the sample, as well as estimate the growth-induced anisotropy parameters. In the following discussion we adopt the frame of reference, where $\varphi_H=$0 corresponds to the geometry, when the projection of the $H_\mathrm{ext}$ on the sample plane is parallel to the $x$-axis (Fig.\,\ref{Fig:Geometry}(b)).

The central goal of our study is to explore the possibility of excitation of the magnetization precession via laser-induced change of the anisotropy parameters (see Eq.(\ref{Eq:Torque})).
Therefore, most of the experiments described below were performed for the field $H_\mathrm{ext}$ directed close to the sample's hard axis ($\zeta_H=80^\mathrm{o}$, $\varphi_H=0$).
In this geometry, when the field makes only a small angle of $10^\mathrm{o}$ with the hard magnetization axis ($x$-axis), a change of any of the anisotropy constants $\Delta K$ is expected to affect the orientation and the value of the effective field $\mathbf{H}_\mathrm{eff}=\mathbf{H}_\mathrm{a}+\mathbf{H}_\mathrm{d}+\mathbf{H}_\mathrm{ext}$, which in turn should result in the finite torque $\mathbf{T}$ according to Eq.(\ref{Eq:Torque}).

\begin{figure*}
\includegraphics[width=16cm]{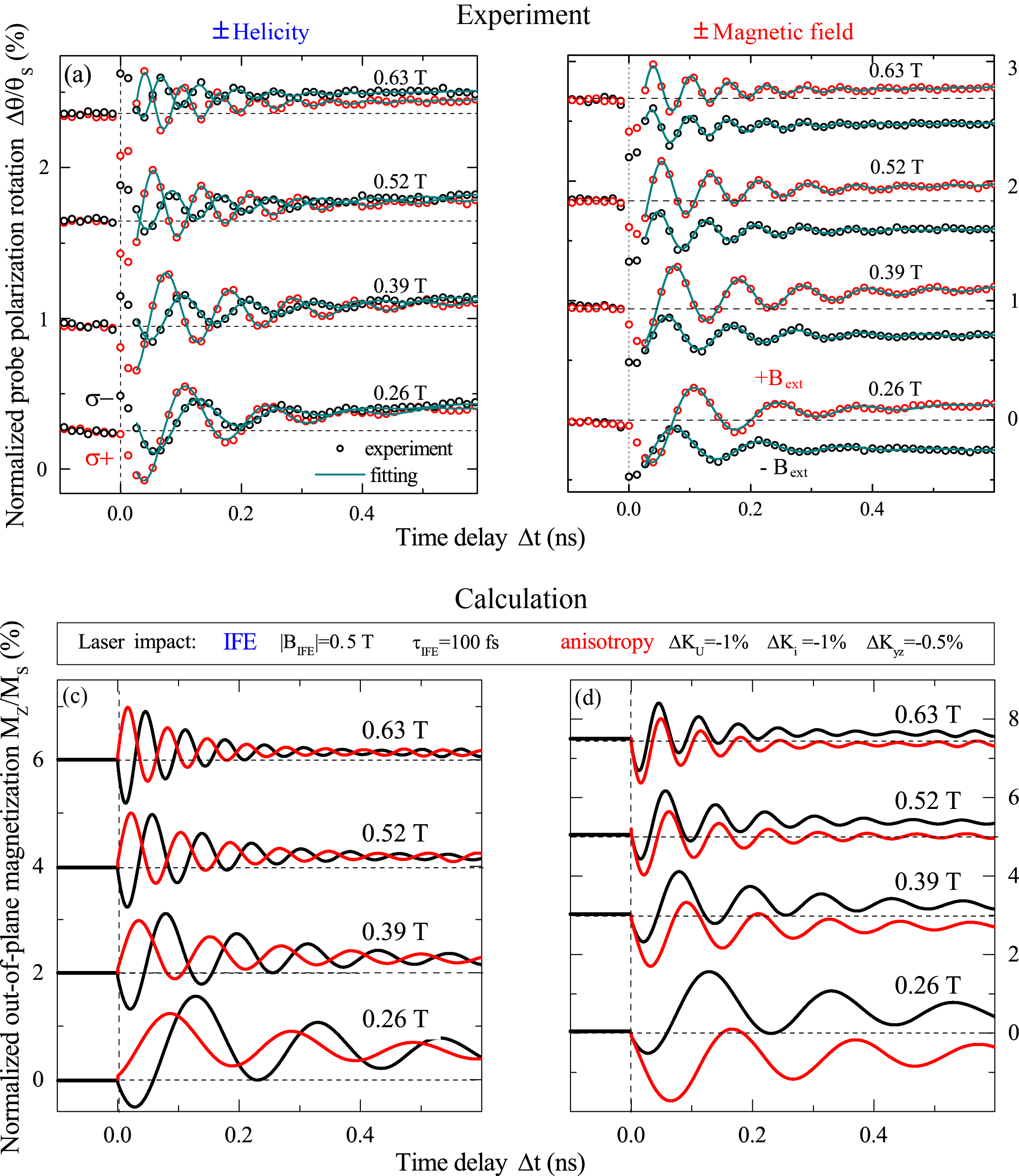}
\caption{(a) Normalized probe polarization rotation induced by the left-($\sigma^\mathrm{-}$) and right-handed ($\sigma^\mathrm{+}$) circularly polarized pump pulses in the MA-film measured as a function of the time delay $t$ for the various magnetic fields. (b) Probe polarization rotation induced by the $\sigma^\mathrm{+}$-polarized pump pulses as a function of the time delay $t$ measured for the various strength of positive and negative magnetic field $\pm H$. (c, d) Simulation of laser-induced Faraday rotation of probe pulse based on Landau–Lifshitz–Gilbert equation for the same cases as (a, b) respectively. Parameters used for this are as follows. $\Delta K_u =  \Delta K_i = -1\,\%; \Delta K_{yz}=-0.5\,\%$; $|B_{\mathrm{IFE}}|=0.5$\,T; $\tau_{\mathrm{IFE}}=100\,$fs}
\label{Fig:FieldDep}
\end{figure*}

\begin{figure}
\includegraphics[width=8.6cm]{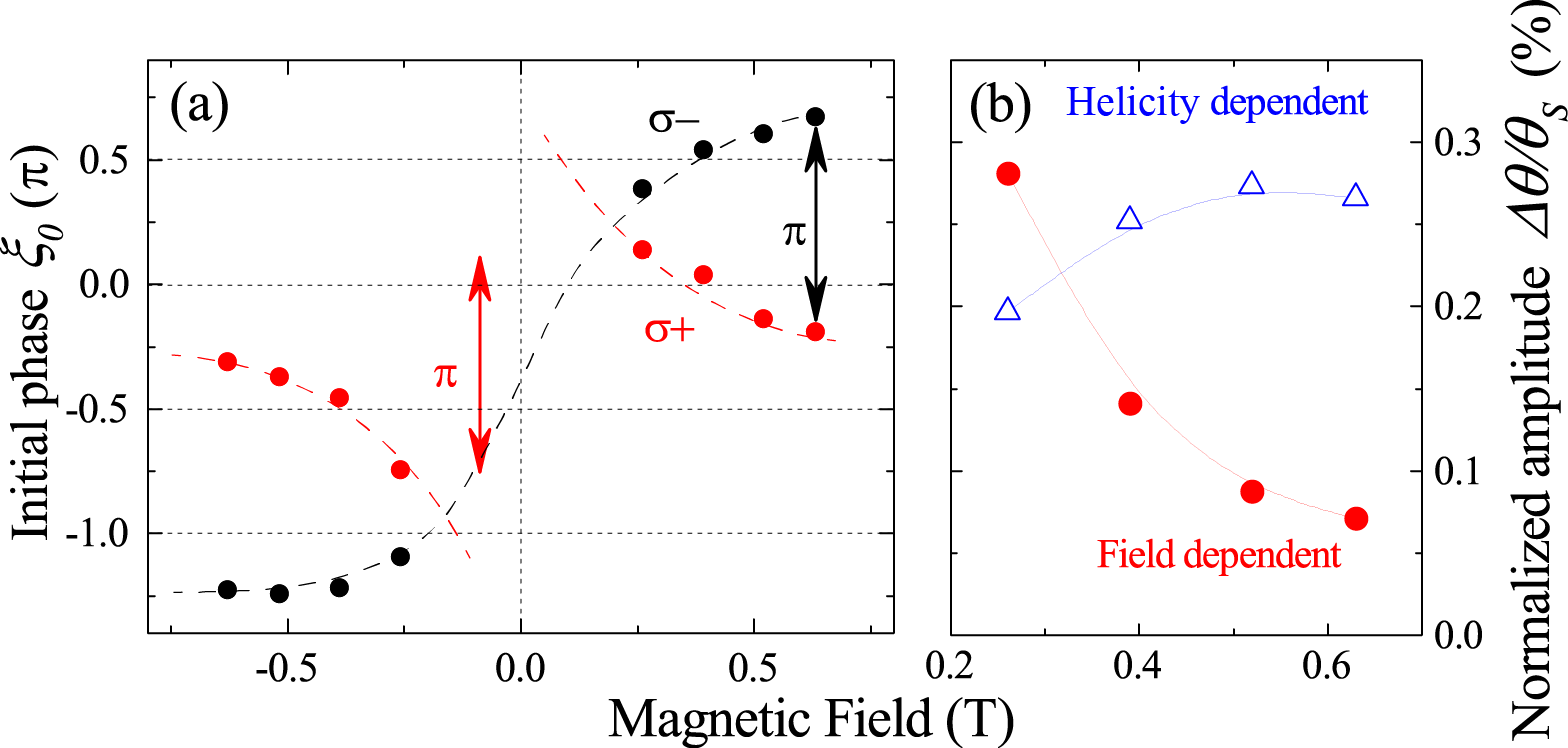}
\caption{(a) Initial phase $\xi_0$ of the probe polarization oscillations excited by $\sigma^\pm$-pump pulses in the MA-film as a function of the applied field, as extracted from the experimental data in Fig.\,\ref{Fig:FieldDep}. (b) Amplitude of the polarization- and field-dependent contribution to the probe polarization oscillations in the MA-film shown in Fig.\,\ref{Fig:FieldDep} as a function of the applied magnetic field strength.}
\label{Fig:FieldDepSummary}
\end{figure}

For revealing possible mechanisms of the laser-induced precession we have studied the influence of the pump pulses polarization and the sign of $H_\mathrm{ext}$ on parameters of the excited dynamics of probe polarization.
In Fig.\,\ref{Fig:FieldDep}(a) we plot the time-resolved dynamics of the probe polarization rotation induced by the right- ($\sigma^\mathrm{+}$) and left-handed ($\sigma^\mathrm{-}$) pump pulses. Fig.\,\ref{Fig:FieldDep}(b) shows the dynamics induced by the $\sigma^\mathrm{-}$-polarized pump pulses measured in the positive and negative applied fields of various strength. One can see that the change of the pump pulse helicity clearly affects the initial phase of the oscillations in a nontrivial way. In order to evaluate the change of the precession parameters we have fitted the experimental curves shown in Fig.\,\ref{Fig:FieldDep}(a) by a function
\begin{equation}
\frac{\Delta\theta(t)}{\theta_s}=\frac{\Delta\theta^0}{\theta_s}e^{-t/\tau_\mathrm{d}}\cos(2\pi ft+\xi_0)+P_2(t),\label{Eq:FitPrecession}
\end{equation}
where $\Delta\theta^\mathrm{0}/\theta_s$ is the normalized oscillations amplitude, $\xi_0$ is the initial phase, $\tau_\mathrm{d}$ is the oscillations decay time, and $P_2(t)$ is the second-order polynomial function accounting for the slowly varying background to be discussed below.
In Fig.\,\ref{Fig:FieldDepSummary}(a) we plot the initial phase of the precession as a function of $H_\mathrm{ext}$.
As one can see, the initial phase $\xi_0$ takes intermediate values, and, furthermore, depends on both the sign and strength of the applied field and the pump polarization. Thus, in the limit of high fields the change of the pump pulse helicity leads to the change of the initial phase of oscillations by $\pi$.
As the applied field decreases this helicity-induced change of the initial phase decreases and almost vanishes in the field of 0.26\,T.
In contrast, when the pump helicity is fixed but the sign of $H_\mathrm{ext}$ is reversed then the phase change by almost $\pi$ occurs at low fields and vanishes at high fields.

Such a complex behavior suggests that there are two competing mechanisms responsible for the precession excitation.
The first one is sensitive to the helicity of the exciting pulse but not to the applied field sign.
The second mechanism is independent from the helicity of the laser pulse but is sensitive to the sign of the applied field.
In the following discussion we refer to these two mechanisms as to the helicity-dependent and field-dependent ones.
The relative contributions of these two mechanisms depend on the applied magnetic field $H_\mathrm{ext}$.
This is demonstrated in Fig.\,\ref{Fig:FieldDepSummary}(b) where we plot the amplitudes of the helicity dependent $\Delta\theta_\mathrm{hd}^0/\theta_s$ and field dependent $\Delta\theta_\mathrm{fd}^0/\theta_s$ contributions to the probe polarization oscillations (Eqs.\,(\ref{Eq:HD},\,\ref{Eq:FD})).

\subsection{Ultrafast inverse Faraday effect}\label{Sec-IFE}

As discussed above the helicity-dependent contribution to the laser-induced magnetization precession becomes larger as the applied field increases.
This indicates that the contribution is stronger when the the equilibrium magnetization orientation becomes closer to the sample plane, and the angle between \textbf{M} and the pump wavevector becomes closer to 90$^\mathrm{o}$ (see Fig.\,\ref{Fig:Geometry}(b)).
At the high-field limit ($H_\mathrm{ext}=0.63$\,T) the initial phase of the oscillations of the probe polarization is changed by a $\pi$ when the helicity of the pump pulse is reversed.
This allows us to draw the conclusion, that the observed excitation of the magnetization precession occurs via the ultrafast inverse Faraday effect (IFE).\cite{Kimel-Nature2005}
This effect was already detected in garnet films.\cite{Hansteen-PRB2006,Jäckl-PRX2017,Savochkin-SciRep2017,Savochkin-SolidState2017}
It microscopically originates from the impulsive stimulated Raman scattering on magnons,\cite{Kalashnikova-PRB2008,Gridnev-PRB2008} and can be described phenomenologically as a femtosecond pulse of an effective magnetic field induced by the circularly polarized laser pulse\cite{Pitaevskii-JETP1961,Pershan-PR1966,Kimel-Nature2005}
\begin{equation}
\mathbf{H}_\mathrm{IFE}\sim\alpha\mathbf{E(\omega)}\times\mathbf{E(\omega)}^*\sim p\alpha I_0\mathbf{z},\label{Eq:IFE}
\end{equation}
where $\mathbf{E(\omega)}$ is the electric field of the light, $\alpha$ is the magneto-optical susceptibility which also defines the Faraday rotation and $I_0$ is the pump intensity. $\mathbf{H}_\mathrm{IFE}$ is directed along the wavevector of the pump pulse, i.e. close to the $z$-axis in our experimental geometry (Fig.\,\ref{Fig:Geometry}(b)).

\begin{table*}
\caption{Components of the torque $\mathbf{T}$ arising due to the ultrafast inverse Faraday effect $\mathbf{H}_\mathrm{IFE}$ and the laser-induced changes of the anisotropy parameters $\Delta K$, as functions of the polar and azimuthal angles $\psi$ and $\varphi_M$, describing the equilibrium orientation of the magnetization.}
\begin{tabular}{c|c|c|c}
  \hline\hline
  \begin{tabular}{c}Driving\\mechanism\end{tabular} & $T_x$ & $T_y$ & $T_z$\\\hline
  $\mathbf{H}_\mathrm{IFE}$ & $\sin\psi\sin\varphi_{M}$ & $-\sin\psi\cos\varphi_{M}$ & 0\\ &&& \\
  $\Delta K_u$ & $-\sin2\psi\sin\varphi_{M}$ & $\sin2\psi\cos\varphi_{M}$ & 0 \\ &&& \\
  $\Delta K_i$ & $\sin2\psi\sin\varphi_{M}$ & 0 & $-\sin^2\psi\sin2\varphi_{M}$ \\ &&& \\
  $\Delta K_{yz}$ & $\cos^2\psi-\sin^2\psi \sin^2\varphi_{M}$ & $\frac{1}{2}\sin^2\psi\sin2\varphi_{M}$ & $-\frac{1}{2}\sin2\psi\cos\varphi_{M}$ \\
  \hline\hline
\end{tabular}\label{Table:Torque}
\end{table*}

In Table\,\ref{Table:Torque} we show the components of the torque related to $\mathbf{H}_\mathrm{IFE}$ as functions of the polar $\psi_\mathrm{M}$ and azimuthal $\varphi_\mathrm{M}$ angles of the magnetization.
These components were derived using Eqs.\,(\ref{Eq:Torque},\ref{Eq:IFE}).
As one can see, the torque increases as the angle $\psi_\mathrm{M}$ between the equilibrium orientation of magnetization and the $z$-axis increases.
In other words, the larger are $M_{x,y}$ components the stronger the created torque (\ref{Eq:Torque}).
This agrees well with the experimental data.
Furthermore, the reversal of the applied magnetic field does not affect the initial phase of $M_{z}$ oscillations excited via IFE\cite{Kimel-Nature2005,Kalashnikova-PRB2008}.
This is also in accordance with our experimental data in the high field limit (see Fig.\,\ref{Fig:FieldDep}(a) and Fig.\,\ref{Fig:FieldDepSummary}(a)). We note, that in Ref.\,\onlinecite{Koene-PRB2015} analogous increase of the precession amplitude excited via IFE with applied field was explained in terms of decreasing demagnetizing field and related ellipticity of the precession trajectory.
In our experimental geometry the observed increase of the precession amplitude is dominated by geometrical effect.

\subsection{Laser-induced change of the magnetic anisotropy}\label{Sec-inducedAnisotropy}

\begin{figure}[h]
\includegraphics[width=8.5cm]{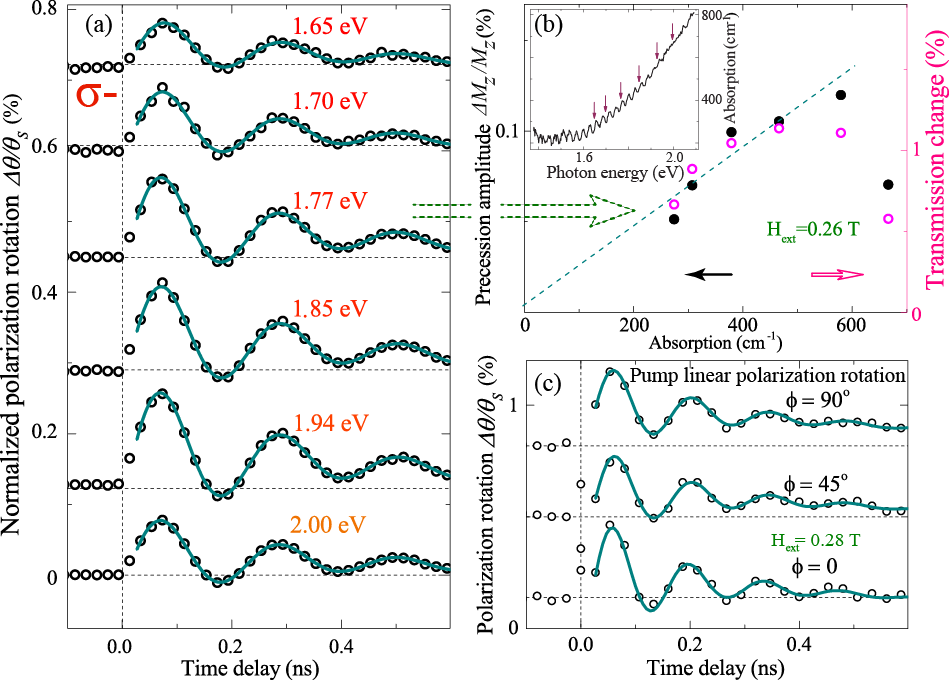}
\caption{(a) Normalized probe polarization rotation induced by the left-handed circularly $\sigma^\mathrm{-}$ polarized pump pulses of different photon energy measured as a function of the time delay $t$ at $H_\mathrm{ext}$=0.26\,T applied along the hard axis. (b) Normalized magnetization precession amplitude and transmission change of the probe beam as a function of absorption coefficient of the pump. Inset depicts absorption coefficient as a function of photon energy of the pump. Apparently oscillations in the inset are connected with the interference of the light in the garnet film. (c) Normalized probe polarization rotation induced by the linearly polarized pump pulses for different slope of pump polarization plane $\phi$ with respect of the probe polarization one measured as a function of the time delay $t$ at $H_\mathrm{ext}$=0.26\,T directed along the hard axis.}
\label{Fig:Absorption}
\end{figure}

As opposed to the inverse Faraday effect discussed in the previous section, the efficiency of the helicity-independent mechanism decreases as the applied field increased.
Previous studies have shown that the linearly polarized femtosecond laser pulse can act as the effective field pulses owing to ultrafast inverse Cotton-Mouton effect, which microscopical nature is similar to that of IFE.\cite{Kalashnikova-PRL2007,Kalashnikova-PRB2008}
Alternatively, linearly-polarized pulses can induce transient anisotropy axis in iron garnets and thus excite the magnetization precession.\cite{Hansteen-PRL2005,Hansteen-PRB2006,Atoneche-PRB2010,Yoshimine-JAP2014,Pashkevich-EPL2014,Koene-PRB2015,Stupakiewicz-arxiv2016}
In order to elucidate possible microscopic mechanism of the laser-induced precession sensitive to the sign of the applied field, we have verified if precession can be excited by the linearly polarized laser pulses, and we have checked whether the azimuthal angle $\phi$ of the pump polarization affects the excitation process.
As shown in Fig.\ref{Fig:Absorption}(c), the precession can be effectively excited by the linearly polarized pulses.
Nevertheless, the initial phase as well as the amplitude of the excited precession are essentially independent from the pump polarization azimuthal angle, which rules out the inverse Cotton-Mouton\cite{Kalashnikova-PRB2008} as well as the photomagnetic effect.\cite{Hansteen-PRL2005}

Polarization-independent excitation of the magnetization precession is often driven by the laser-induced demagnetization which affects $\mathbf{H}_\mathrm{eff}$ via modification of $\mathbf{H}_\mathrm{d}$ (\ref{Eq:Heff}).
This mechanism can be safely ruled out in the considered experiment, since the laser-induced demagnetization in the studied garnet is the slow process with the characteristic time of $\sim$500\,ps (see Appendix\,\ref{App-demagnetization}).
Alternatively, the observed precession excitation could occur via fast laser-induced changes of the magnetic anisotropy parameters.

\begin{figure}[h]
\includegraphics[width=8.5cm]{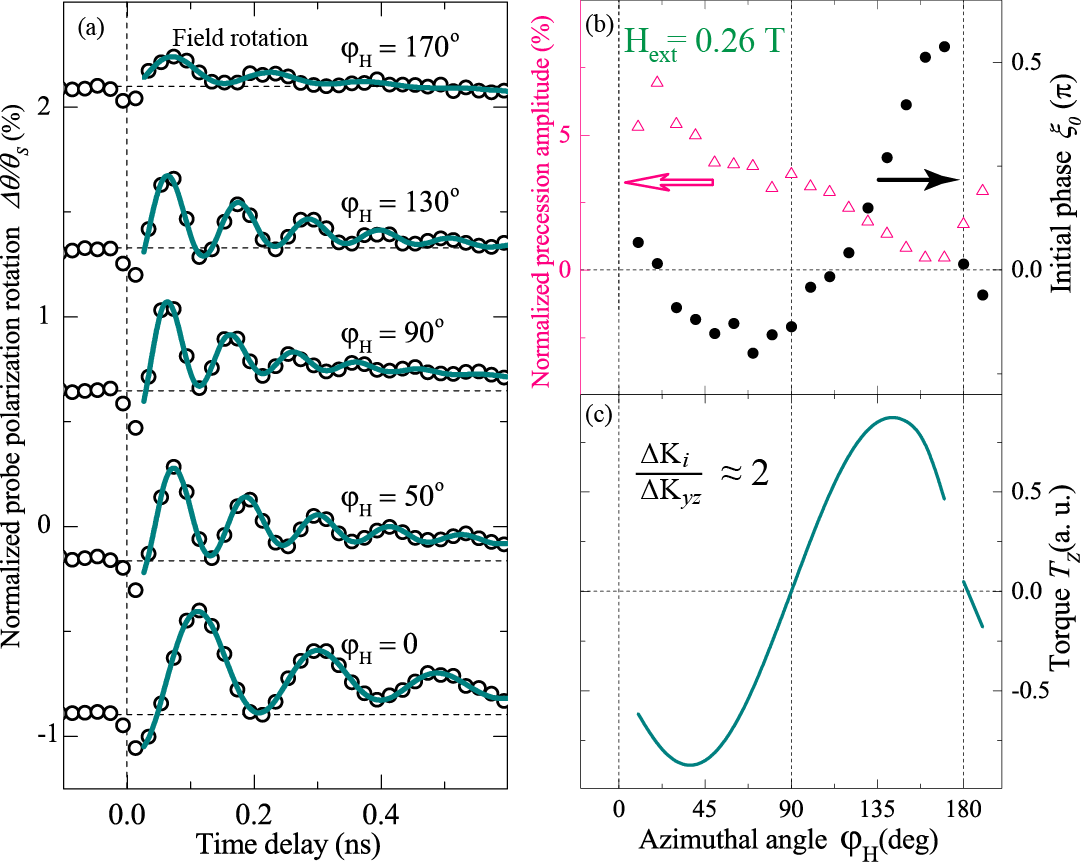}
\caption{(a) Normalized probe polarization rotation induced by the left-circularly $\sigma^\mathrm{-}$ polarized pump pulses of different azimuthal angles $\varphi$ measured as a function of the time delay $t$. (b) Initial phase $\xi_0$ and amplitude of the laser-induced precession as a a function of $\varphi_{H}$ measured for external magnetic field of $H_\mathrm{ext}$=0.26\,T in the MA-film. (c) $T_z$ component of the torque (\ref{Eq:Torque}) calculated as a function of $\varphi_{H}$ for a case when $\Delta K_{i}$ is twice as large as $\Delta K_{yz}$. Note, that the abrupt change of the initial phase $\xi_0$ observed at $\varphi_{H}=180^{\mathrm{o}}$ occurs because the effective anisotropy field $H_{a}$ jumps to another equilibrium position when the $\varphi_H>180^\mathrm{o}$.}
\label{Fig:Phases}
\end{figure}

In order to test this hypothesis 
we have studied in more details the laser-induced excitation of the precession in the magnetic field of $H_\mathrm{ext}=$0.26\,T applied at different azimuthal angles in the range $\mathrm{10}^\mathrm{o}<\varphi_H<\mathrm{190}^\mathrm{o}$  (Fig.\,\ref{Fig:Phases}(a)).\cite{Ma-JAP2015}
Applied field of 0.26\,T has been chosen in order to maximize the contribution to the precession excitation originating from the field-dependent mechanism.
Fig.\,\ref{Fig:Phases}(b) shows the amplitude $\Delta M_z/M_s$ and the initial phase $\xi_0$ of the precession as a function of the applied field azimuthal angle $\varphi_{H}$.
As one can see, $\xi_0$ is gradually changing by more than $\pi/2$ in the range $\mathrm{80}^\mathrm{o}<\varphi_H<\mathrm{170}^\mathrm{o}$. This experiment clearly demonstrates that the initial phase of the precession is very sensitive to the orientation of the applied magnetic field, and, thus, to the equilibrium orientation of the magnetization with respect to the sample axes.

We note that the change of the anisotropy parameters can be considered as a displacive mechanism of the excitation of the precession when the spin system reacts to the changes of $H_\mathrm{eff}$ direction under the impact of the laser pulse.\cite{Kalashnikova-PRB2008} In this case the phase $\xi_0\,=0$ of the oscillations of the $M_z$-component of the magnetization (\textit{cosine}-like temporal dependence) indicates that the corresponding initial torque vector is directed in the $xy$ plane, i.e. the $z$ component of $\mathbf{H}_\mathrm{eff}$ has been altered.
The phase $\xi_0=\pi/2$ corresponds to the situation, when the excitation leads to the change of the component of the effective field in the $xy$ plane.
Consequently, the laser-induced torque acting on the magnetization is directed along the $z$-axis. Thus, it can be seen from Fig. \ref{Fig:Phases}(a,b) that the torque changes its direction and strength.

In the Table\,\ref{Table:Torque} we present the components of the torque (\ref{Eq:Torque}) as a function of the polar $\psi$ and azimuthal $\varphi_\mathrm{M}$ angles of the magnetization.
These expressions allow us to eliminate the change of the $K_u$ parameter as a major driving mechanism for the precession excitation.
Indeed, for any orientation of the magnetization the no $T_z$-component related to the $\Delta K_u$ occurs.
In this case the initial phase $\xi_0$ of the $M_z$-oscillations is expected to be 0 with no pronounced dependence on $\varphi_H$.
This strongly contradicts the experimental data in Fig.\,\ref{Fig:Phases}(b), where the initial phase $\xi_0$ spans over a range exceeding $\pi/2$, thus indicating that at least for some directions of the applied magnetic field the $T_z$-component is nonzero.

Another important conclusion which can be drawn from the results shown in Fig.\ref{Fig:Phases}(b) is that the initial phase $\xi_0$ is close to zero when the azimuthal angle is $\varphi_{H}$=90$^\mathrm{o}$.
Note, that in this case both the effective anisotropy field $\mathbf{H}_\mathrm{a}$ and the applied magnetic field lie in the $yz$ plane (see Fig.\,\ref{Fig:Geometry}(a)), which is the easy plane for the magnetization.
In this geometry, the net effective field $\mathbf{H}_\mathrm{eff}$ should remain in the $yz$ plane even if the laser-induced changes of any of the anisotropy parameters $\Delta K$ occur.
This, in turn, corresponds to the initial phase $\xi_0=0$ of the laser-induce precession, in agreement with that observed in the experiment.

As one can see from the data presented in Fig.\,\ref{Fig:Phases}(a), there is a clear decrease of the amplitude of the excited precession as the $\varphi_H$ is changed from 10$^\mathrm{o}$ to 170$^\mathrm{o}$.
It can be easily understood, taking into account that in this case the azimuthal angle of magnetization varies from a value $10^\mathrm{o}<\varphi_M<90^\mathrm{o}$ to $90^\mathrm{o}<\varphi_M<170^\mathrm{o}$.
The polar angle of magnetization varies as well, remaining in the range of $0^\mathrm{o}<\psi<90^\mathrm{o}$.
Then, from the expressions listed in Table\,\ref{Table:Torque} one can see, that such change of the magnetization directional angles results in the sign change of some, but not all components of the torque, excreted due to laser-induced change.

\begin{figure*}[h]
\includegraphics[width=14cm]{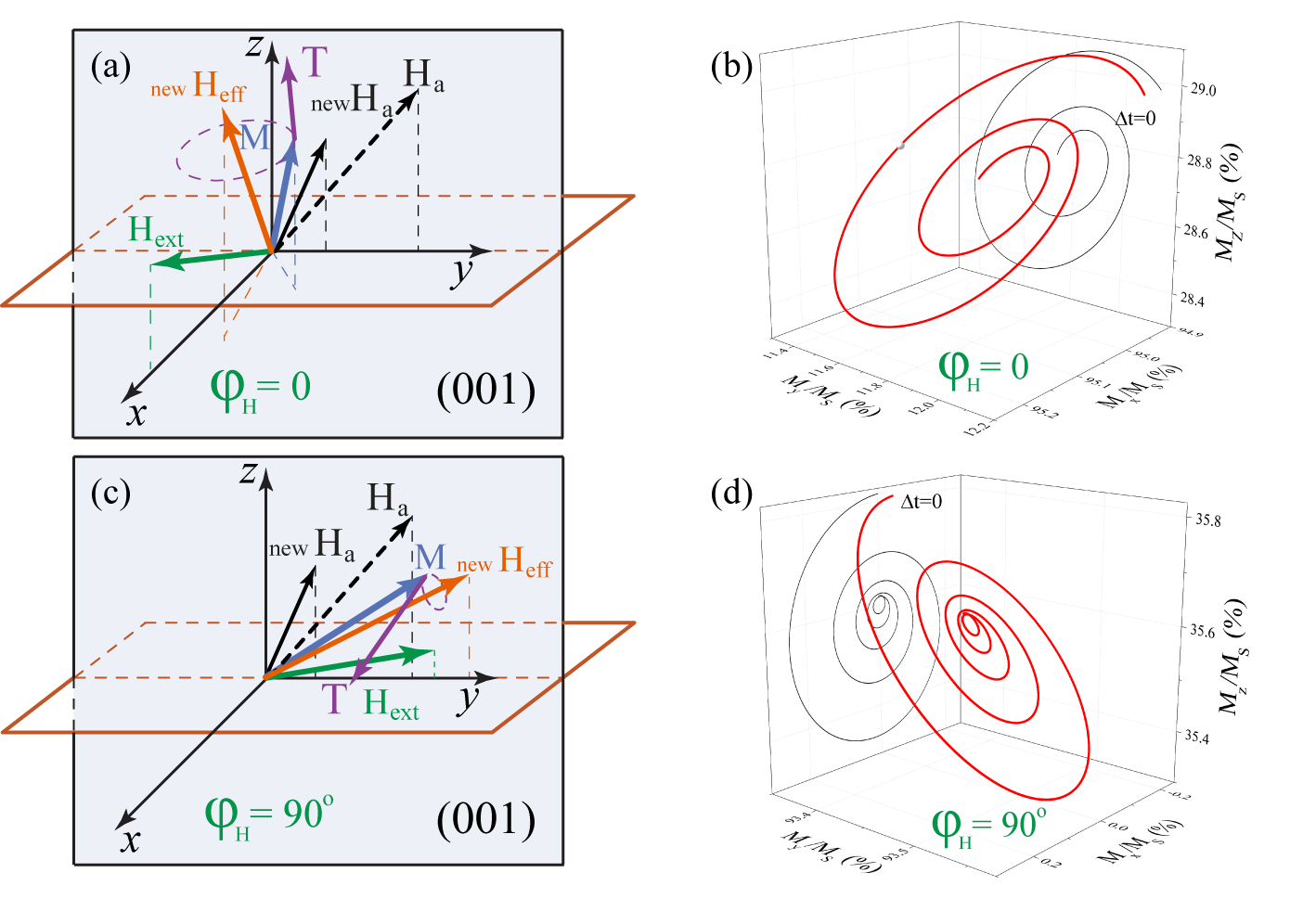}
\caption{Graphical illustration of the process of pulse-induced magnetic anisotropy change with the following precessional dynamics for (a, b)\,$\varphi_{H} = 0$  and (c, d)\,$\varphi_{H} = 90^\mathrm{o}$. Cases (b) and (d) depict modeling of the magnetization trajectory after the excitation using parameters as in Fig.\ref{Fig:FieldDep}. Note, that in the cases (c, d), the effective anisotropy field and the applied field are both lie in the ($yz$)-plane. As a result, induced change of the anisotropy parameters does not deflect the net effective field away from this plane. This prevents appearance of the $T_z$-component of the induced torque. By contrast, when the applied field is perpendicular to the $yz$ plane, the same change of the anisotropy modifies both deflection of $\mathbf{H}_\mathrm{eff}$ from the $yz$ plane and its orientation in the plane. As a result, all three components of the induced torque $\mathbf{T}$ should be finite.}
\label{Fig:Mechanism}
\end{figure*}

In order to further substantiate suggested mechanism of the precession excitation, we have modeled the changes of the different components of the torque (\ref{Eq:Torque}) as functions of either $K_i$ or $K_{yz}$ modified under the action of the pump pulse.
The calculations were performed taking anisotropy parameters obtained for this film.
The equilibrium orientation of the magnetization ($\psi$, $\varphi_M$) has been calculated using these parameters for each direction of the $H_\mathrm{ext}$ in the range of $\mathrm{10}^\mathrm{o}<\varphi_H<\mathrm{190}^\mathrm{o}$.
Then the laser-induced torque (\ref{Eq:Torque}) has been calculated assuming changes of $K_i$ or $K_{yz}$ anisotropy parameters.
For the sake of clarity the torque occurring due to the ultrafast inverse Faraday effect was neglected.
Fig.\,\ref{Fig:Phases}(c) shows $T_z$-component of the torque (\ref{Eq:Torque}) as a function of $\varphi_{H}$ for the case when the laser-induced decrease of the orthorhombic anisotropy parameter $\Delta K_{yz}$ is twice as large as the change of the in-plane uniaxial parameter $\Delta K_{i}$.
As one can see, the outcome of this model agrees with the experimental results.
In fact, the variations of the calculated $T_z$-component of the torque (Fig.\,\ref{Fig:Phases}(c)) correlate with the changes of the initial phase of the precession $\xi_0$ (Fig.\,\ref{Fig:Phases}(b)). In Fig.\,\ref{Fig:Mechanism}(c) we show in details the calculated time-dependent trajectory of the magnetization upon sudden change of naisotropy parameters in two distinct situations, $\varphi_H$=0 and $\varphi_H$=90$^\mathrm{o}$. Obtained trajectories agree with the experimental results and reproduce the $cosine$-like variations of $M_z$ when $\varphi_H$=90$^\mathrm{o}$. We note that although the $T_z$-component of the torque vanishes for certain angles $\varphi_H$ the total torque remains always nonzero. This is supported by the experimental observations, that for any azimuthal angle in the range $\mathrm{10}^\mathrm{o}<\varphi_H<\mathrm{190}^\mathrm{o}$ the magnetization precession is still excited. In Fig.\,\ref{Fig:FieldDep}(c,d) we show the change of $M_z$ as a function of the time delay, calculated using LLG equation and taking into account both the ultrafast IFE and the change of anisotropy parameters. As can be seen, we successfully reproduce the field and polarization dependences (Fig.\,\ref{Fig:FieldDep}(a,b)).

The described above analysis allows us to conclude that the field-dependent excitation of the magnetization precession occurs via the ultrafast change of the growth-induced anisotropy parameters.
Since this change is independent from the laser pulse linear polarization (Fig.\,\ref{Fig:Absorption}\,(c)), we argue that the heating can be a plausible mechanism underlying the process.
Indeed, various parameters of the growth-induced anisotropy are known to be temperature dependent.\cite{Shumate-JAP1973, Zener-PR1954}
Thus, the increase of the lattice temperature in magnetic dielectrics on a picosecond time scale following the absorption of the fraction of the femtosecond laser pulse energy, should affect the magnetic anisotropy on the same time scale.
The growth-induced anisotropy parameters exhibit distinct temperature dependencies, and, as a result, they change differently in response to the heating.
The subsequent cooling of the lattice relies on the much slower heat dissipation.
Consequently, the relaxation of the effective anisotropy field to its equilibrium value is expected to be the slow process, in agreement with the displacing character of the change of the anisotropy parameters.

For the pump fluence of 10\,mJ/cm$^{-2}$ we estimated the lattice temperature increase to be of $\Delta T\sim$1\,K, taking the heat capacity of a Bi-substituted iron garnet of 3.85\,J$\cdot\mathrm{cm}^{-3}\cdot\mathrm{K}^{-1}$.\cite{Inoue-JJAP1980} To verify if such small $\mathrm{\Delta T}$ could lead to the anisotropy change used in our calculations, we used the phenomenological relation between the temperature dependent magnetization and anisotropy in a medium with uniaxial anisotropy:\cite{Zener-PR1954}
\begin{equation}
1+\frac{\Delta K_u(T)}{K^{0}_{u}} = 1+\left(\frac{\Delta M(T)}{M^{0}}\right)^3, \label{Eq:KvsM}
\end{equation}
where $K^{0}_{u}$ and $M^{0}$ is anisotropy parameter and magnetization at the temperature $T$ = 0, respectively.
In our sample demagnetization amounts to $\Delta M \approx$ -0.2\,\% (see App.\,\ref{App-demagnetization}) and can be used as a measure of total $\mathrm{\Delta T}$ occured due to laser pulse excitation. Then Eq.(\ref{Eq:KvsM}) yields that the anisotropy change for the same increase of the temperature is of $\Delta K_u\approx$ -0.6\,\%. This is of the same order of magnitude that we used in our modeling (Fig.\,\ref{Fig:FieldDep}(c,d)).

In order to confirm the conclusion about the thermal origin of the laser-induced anisotropy change, we made a series of experiments with different pump photon energy. This allowed us to study the precession excitation when the absorption coefficient for the pump pulses varies (see inset in Fig.\,\ref{Fig:Absorption}(b)). The latter leads to different levels of the ultrafast lattice heating achieved at the same excitation energy. Hence we expect the increase of the precession amplitude as the pump photon  energy rises. Fig.\,\ref{Fig:Absorption}\,(a) shows pump-induced normalized probe polarization rotation as a function of time delay $t$ at $H_\mathrm{ext}$=0.26\,T for different pump photon energy $E_\mathrm{ph}$. Precession amplitudes presented in Fig.\,\ref{Fig:Absorption}\,(b) are extracted from fit of the experimental data to the function (\ref{Eq:FitPrecession}). The precession amplitude rises with absorption, as expected, with one exception of the highest absorption at $E_\mathrm{ph}$=2\,eV. In our opinion this deviation from a monothonous increase may originate from nonuniform excitation due to the high absorption. Similar deviation from monotonously increasing trend of the precession amplitude dependence on the pump pulse absorbtion was also observed in the Ref.\,\onlinecite{Savochkin-SolidState2017}, and was ascribed to generation of magnetostatic spin waves. We note also that, in our experiments optical transmission change also gradually increases with increase of absorption, except of the case of the highest absorption coefficient, indicating that some optical nonlinear effects may start to play a role in this absorption levels.

We note that the polarization-independent excitation of the magnetization precession in garnets has been also reported in Refs.\,\onlinecite{Hansteen-PRB2006},\,\onlinecite{Koene-PRB2015}. Both studies were carried out on the films grown on high-symmetry (001) GGG substrates and possessing easy-plane-type anisotropy, with no pronounced in-plane anisotropy. In Ref.\,\onlinecite{Hansteen-PRB2006} it has been shown that the polarization independent contribution to the precession excitation occurs with magnetic field being deflected from the sample plane.
The authors have interpreted the mechanism of the precession excitation as the nonthermal photo-induced anisotropy.
Nevertheless, we suggest that the thermal change of the anisotropy could also contribute to the excitation process.
However, distinguishing this mechanism from the others would be challenging, since the (001) films do not allow for experimental studies, where the torque occurring due to the laser-induced thermal changes of magnetic anisotropy possesses strong dependence on the applied field orientation (Fig.\,\ref{Fig:Phases}).

The demagnetization induced in dielectric garnets by pump pulse, which is also a result of the heating, is mediated by the relatively slow phonon-magnon interaction. As we show (see App.\,\ref{App-demagnetization} and Fig.\,\ref{Fig:Demagnetization}(b) therein) the demagnetization in the garnet films upon optical excitation in a range of moderate absorption occurs on the time scale of $\sim$500\,ps. Therefore, the demagnetization itself cannot contribute to the excitation of the precession, however, it does contribute to the observed probe polarization dynamics. In our experimental geometry the static $M_z$-component is finite in the whole range of the applied fields (see Fig.\,\ref{Fig:Faraday}(a)). Therefore, the demagnetization in the sample under consideration, analogous to that observed in the film with much stronger anisotropy, should manifest itself as a slow change in the rotation of the probe polarization.
Indeed, as one can see form Fig.\,\ref{Fig:FieldDep}(b), there is a slow change of the induced probe polarization with a characteristic time of $\sim$700\,ps.
This slow process depends on the sign of the applied field, does not depend on the pump polarization, and is stronger in the range of low magnetic field.
This contribution to the signal is the manifestation of the laser-induced demagnetization. The character of the time delay dependence of the $z$-component of the magnetization is somewhat more intricate than the exponential decay, found in the film with strong anisotropy (Fig.\,\ref{Fig:Demagnetization}(b)).
Possibly, the change of the magnetization value, which follows the function (\ref{Eq:FitDemagn}), is accompanied by the change of the effective anisotropy field (\ref{Eq:Heff}). As a result, the change of the $M_z$ can deviate from the exponential behaviour (\ref{Eq:FitDemagn}).

\section{Conclusions}\label{Sec-conclusion}

In conclusions, we have investigated the interaction between femtosecond laser pulses and thin ferrimagnetic substituted iron garnet film grown on a low-symmetry (210)-oriented GGG substrate.
We show experimentally that the impact of a laser pulse results in magnetization precession.
Using pump pulses with various polarizations we demonstrated that the precession is in fact excited via two distinct mechanisms.
The first one is the ultrafast inverse Faraday effect, which is found by now in a large number of magnetic dielectrics.
Competing with this mechanism, is the change of the growth-induced anisotropy.
Lack of the pump polarization dependence, as well as slow relaxation time of this mechanism indicate that there is a thermal change of the magnetic anisotropy, triggered by rapid increase of the lattice temperature.
We note, that, in contrast to laser-induced dynamics in magnetic metals, this excitation does not rely on the much slower demagnetization.

The excitation of the magnetization precession in magnetic dielectrics via thermal change of the anisotropy has been previously explored only in the vicinity of the orientation phase transitions, where the anisotropy is strongly temperature dependent.
Here we show that even far from the phase transition region the ultrafast heating of the lattice, resulting in the anisotropy change, can effectively excite the magnetization precession.
Interestingly, the amplitude of the precession, excited via this mechanism is comparable to that, occurring due to the IFE, which is expected to be pronounced in the studied garnet films.
Furthermore, the relative contributions from these mechanisms can be changed by varying the applied magnetic field.
As a result, one can gradually control the initial phase of the precession and its sensitivity to the polarization of the exciting laser pulse.

As we demonstrated experimentally and phenomenologically, the parameters of the magnetization precession excited via ultrafast change of the anisotropy are sensitive to the orientation of the applied magnetic field.
Most importantly, the initial phase of the precession changes drastically, depending on the angle of the field with the easy plane of the magnetization.
This suggests, that the ultrafast change of the magnetic anisotropy should always be considered when one deals with the precession excited by femtosecond laser pulses in a dielectric placed in an external magnetic field not collinear with the easy magnetization direction.

\section{Acknowledgements}

This work was performed at Ferroics Physics Laboratory (Grant No. 14.B25.31.0025). Experiments were performed under support of Russian Foundation for Basic Research (grants No.\,15-02-09052-a and 16-02-00377-a). Analytical work and spectral studies were performed by L.A.Sh., P.Yu.Sh, and A.M.K. under the support of the Russian Science Foundation (grant No.\,16-12-10485). We thank, M. P. Scheglov for verification of the samples orientation, Dr. M. V. Zamoryanskaya for XRF characterization, Dr. M. P. Volkov for the magnetic characterization using the system PPMS, and Dr. L. V. Lutsev for performing FMR measurements.

\appendix
\section{Equilibrium growth- and stress-induced anisotropy of the (210)-film}\label{App-anisotropy}

Magnetic anisotropy of the garnet film grown on a substrate of arbitrary orientation and characterized by a pronounced growth- and/or stress-induced anisotropy can be found from the two-parameter model:\cite{Gyorgy-APL1971,Eschenfelder-book}
\begin{eqnarray}
w_a&=&A(\alpha_{x'}^2\beta_{x'}^2+\alpha_{y'}^2\beta_{y'}^2+\alpha_{z'}^2\beta_{z'}^2)\label{Eq:appA:nisotropy}\\
&+&B(\alpha_{x'}\alpha_{y'}\beta_{x'}\beta_{y'}+\alpha_{y'}\alpha_{z'}\beta_{y'}\beta_{z'}+\alpha_{z'}\alpha_{x'}\beta_{z'}\beta_{x'})\label{Eq:app:Anisotropy},\nonumber
\end{eqnarray}
where $\alpha_i$ and $\beta_i$ are directional cosines of magnetization \textbf{M} and the sample normal with respect to the crystallographic axes $x',\,y'\,z'$.
By taking into account the growth direction [210] and making the transformation from the crystallographic axes to the coordinate frame $xyz$ shown in Fig.\,\ref{Fig:Geometry}, one finds the expression for the anisotropy parameters $K$ entering Eq.\,(\ref{Eq:anisotropy}):
\begin{eqnarray}
K_u&\approx&0.67A+0.16B;\nonumber\\
K_i&\approx&0.67A-0.16B;\label{Eq:app:constants}\\
K_{yz}&\approx&2.13K_u-0.63K_i.\nonumber
\end{eqnarray}
In order to estimate the values of the anisotropy parameters we have approximated the experimental field and azimuthal dependencies of the precessional frequency (Fig.\,\ref{Fig:Precession}\,(b,c)), and the field dependence of the Faraday rotation $\theta_F\sim M_z$ (Fig.\,\ref{Fig:Faraday}\,(a)) using the analytical expression for magnetic energy.\cite{Suhl-PR1955,Artman-PR1957} The resulting curves are shown in the corresponding plots.
The values of the anisotropy parameters providing good agreement between calculations and experimental data are given in Table.\,\ref{Table:parameters}.

The magnetic anisotropy described by the parameters $K$ has two contributions, the growth- ($K^g$) and the stress-induced ($K^s$) ones.
Straightforward calculation of the growth-induced contribution to the magnetic anisotropy requires knowledge of a number of microscopical parameters, while the stress-induced one can be estimated from the phenomenological model.
Magnetoelastic contributions to the magnetic anisotropy energy of the (210) iron garnet film can be expressed as\cite{Eschenfelder-book}
\begin{eqnarray}
K_u^s&=&\frac{3}{2}\lambda_{100}\sigma_0\beta_1^2+3\lambda_{111}\sigma_0\beta_1^3\beta_2;\nonumber\\
K_i^s&=&\frac{3}{2}\lambda_{100}\sigma_0\beta_2^2+3\lambda_{111}\sigma_0\beta_1\beta_2^3;\\
K_{yz}^s&=&-3\lambda_{100}\sigma_0\beta_1\beta_2,\nonumber
\label{Eq:ME}
\end{eqnarray}
where$\lambda_{100}$ and $\lambda_{111}$ are magnetostriction coefficients for the iron garnet film, and $\sigma_0$ is the biaxial stress arising due to the film/substrate lattice mismatch $\Delta a_0/a_0$.
For garnets, characterized by the Young's modulus of $\sim2\cdot10^{11}$\,N/m$^2$ the relation between the stress and the substrate/lattice mismatch $\Delta a_0/a_0$ is $\sigma_0=2.8\cdot10^{11}\Delta a_0/a_0$\,Pa.\cite{Davies-JMS1975}
We estimated the stress-induced contribution to the magnetic anisotropy, taking the lattice mismatch in the studied film $\Delta a_0/a_0=-0.38$\,\% and the typical magnetostriction coefficients for the garnets $\sim-10^{-6}$.
Resulting stress-induced anisotropy constants are given in Table\,\ref{Table:parameters}.

\begin{table}
\caption{Total, growth-, and stress-induced anisotropy parameters of the studied film, as extracted from the experimental data and calculations.}
\begin{tabular}{l|c|c|c}
  \hline\hline
  & Total ($K$) & Growth ($K^g$) & Stress ($K^s$)\\\hline
  $K_u$\,(J/m$^3$) & $-5\cdot10^{3}$ & $-7\cdot10^3$ & $2\cdot10^3$\\
  $K_i$\,(J/m$^3$) & $-3\cdot10^{3}$ & $-3.5\cdot10^3$ & $5\cdot10^2$ \\
  $K_{yz}$\,(J/m$^3$) & $-8.7\cdot10^{3}$ & $-7.7\cdot10^3$ & $-1\cdot10^3$\\
  \hline\hline
\end{tabular}\label{Table:parameters}
\end{table}

The values of the growth-induced anisotropy parameters calculated as $K^g=K-K^s$ are listed in Table\,\ref{Table:parameters}.
From comparison of the the growth- and stress-induced parameters we can conclude that the growth-induced anisotropy dominates over the stress-induced one.
This is in agreement with the literature data on substituted garnets with high ($>1$) Bi$^{3+}$ content, in which growth-induced anisotropy energy can exceed $10^4$\,J/m$^3$.\cite{Hansen-JAP1985,Eschenfelder-book}

\section{Laser-induced demagnetization of the substituted iron garnet film}\label{App-demagnetization}

\begin{figure}
\includegraphics[width=8.5cm]{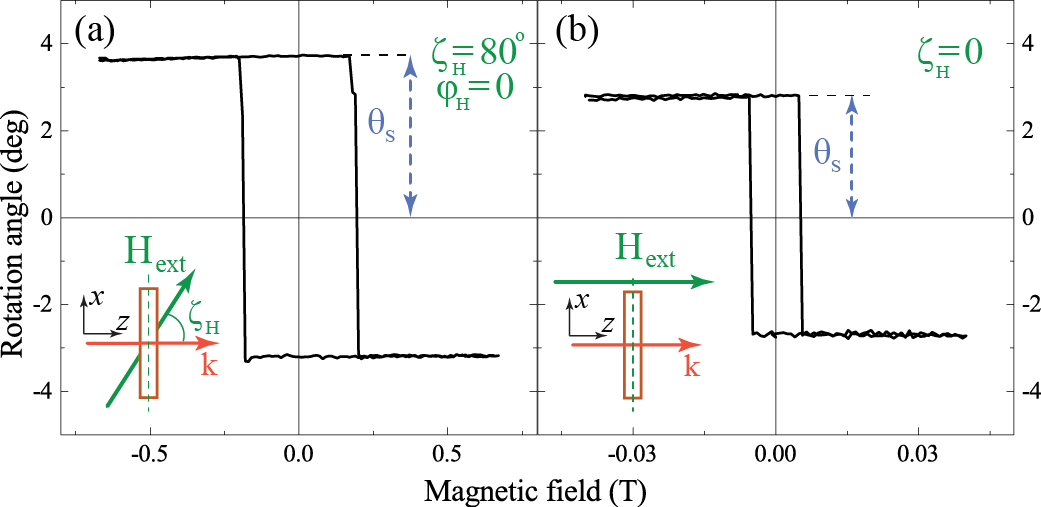}
\caption{(Color online) Static Faraday rotation as a function of the external magnetic ﬁeld in the SA-film with external field applied at (a) $\psi_\mathrm{H} = 80^\mathrm{o}$ and (b) $\psi_\mathrm{H} = 0$. $\theta_s$ denotes the Faraday rotation at remanence, which is proportional to the samples magnetization $M_S$. Insets show the experimental geometries used for measuring the static Faraday rotation in the samples.}
\label{Fig:Faraday-1}
\end{figure}

Laser-induced demagnetization, if occurs on a relatively short time scale, can be a driving mechanism of the excitation of the magnetization precession.
Therefore, we investigated the demagnetization triggered by the femtosecond laser pulses in order to evaluate the strength and the time scale of this process in the garnet film under study.
For this we selected the film of composition (Y$_{1.17}$Bi$_{1.44}$Pr$_{0.26}$Lu$_{0.24}$)(Fe$_{3.66}$Ga$_{1.23}$)O$_{12}$, which is characterized by the strong out-of-plane anisotropy.
Fig.\,\ref{Fig:Faraday-1} shows the field dependencies of the Faraday rotation measured in the same geometries as in Fig.\,\ref{Fig:Faraday}.
As one can see, available external field applied at the $\mathrm{80^o}$ with respect to the sample normal is not sufficient to deflect the magnetization from the easy axis (Fig.\,\ref{Fig:Faraday-1}(a)).
In such film in the experimental geometry, used for the pump-probe studies (Fig.\,\ref{Fig:Geometry}(b)) with the field applied at $\zeta_H = 80^0$, we probe solely the change of the magnetization magnitude while other processes such as magnetization precession are not detected even if excited.
Therefore, such experiment is the most suitable for detecting laser-induced demagnetization via the Faraday rotation.

\begin{figure}
\includegraphics[width=8.5cm]{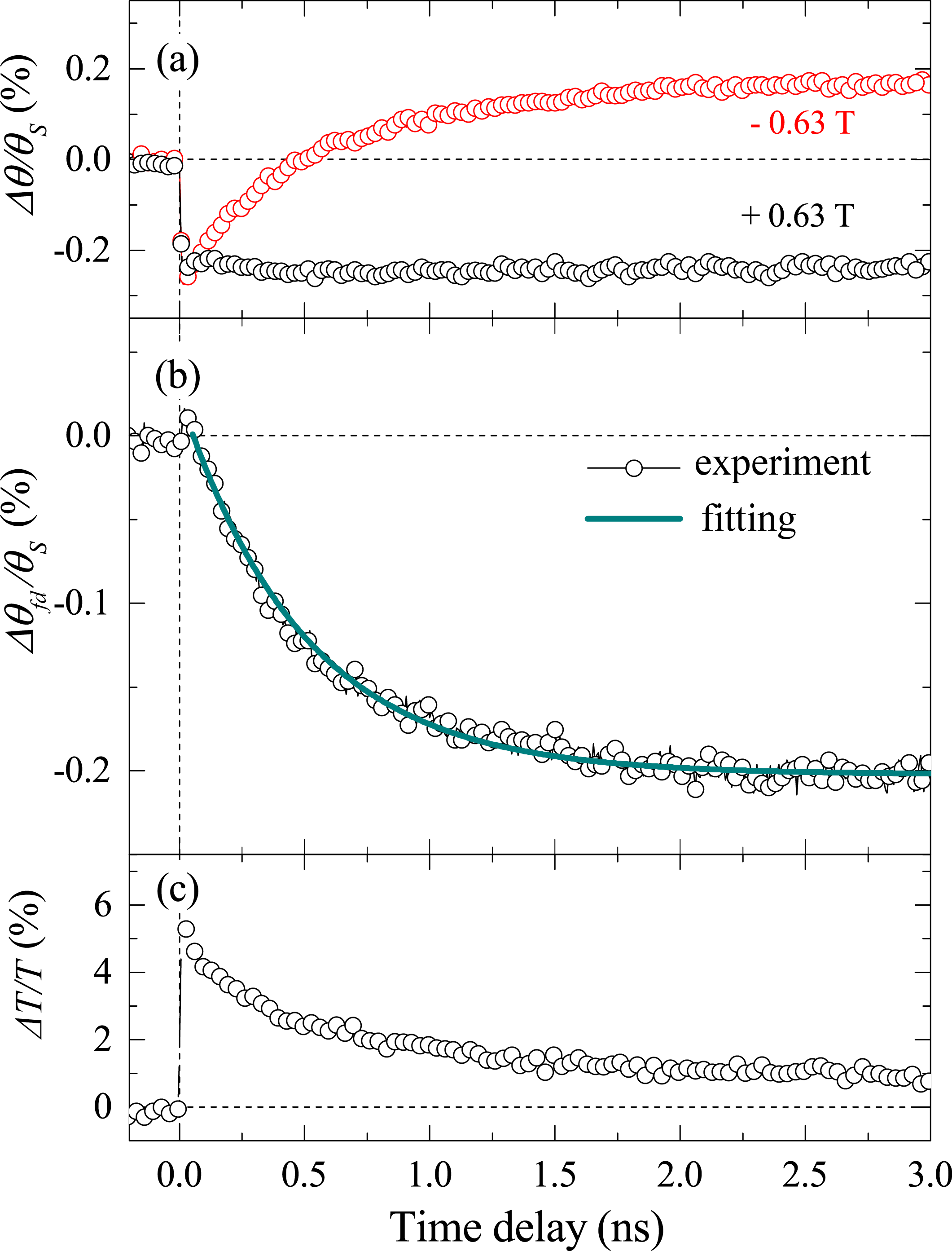}
\caption{(Color online) (a) Normalized rotation of the probe polarization induced by the linearly polarized laser pulse as a function of the pump-probe time delay $t$ measured for positive and negative magnetic field of $\pm$ 0.63\,T in the SA-film. (b) Field-dependent contribution to the laser-induced dynamics, extracted using Eq. (\ref{Eq:FD}) from the data shown in the panel (a). Solid line is a fit using the Eq.\,(\ref{Eq:FitDemagn}). (c) Normalized change of the sample transmission as a function of pump-probe time\,delay\,$t$.}
\label{Fig:Demagnetization}
\end{figure}

Dependencies of the polarization rotation of the probe pulse as a function of the pump-probe time delay $t$ are shown in Fig.\,\ref{Fig:Demagnetization}(a) for the two opposite applied magnetic fields exceeding the saturation field.
Fig.\,\ref{Fig:Demagnetization}(a) shows that the dynamics notably changes as the $\mathbf{H}_\mathrm{ext}$ is reversed.
In order to analyse this behavior we extracted the sign-dependent contribution (\ref{Eq:FD}) from the measured signals, which can be reliably considered as a measure of the pump-induced change of $M_z$.
We note that such approach excludes the sign-independent contribution.
Time evolution of the latter (not shown) closely resembles the temporal transmission changes $\Delta T/T$ caused by pump shown in Fig.\,\ref{Fig:Demagnetization}(c).
Therefore, we attribute this $\Delta\theta(t;\pm H)$ contribution to the changes of optical or magneto-optical sample properties but not to the alteration of the magnetization magnitude or its orientation.

The time-delay dependence of the signal $\Delta\theta_\mathrm{fd}(t)/\theta_s\sim\Delta M_z(t)/M_s$ has been fitted by a function
\begin{equation}
\frac{\Delta\theta_\mathrm{fd}(t)}{\theta_s}=\frac{\theta_\mathrm{dm}}{\theta_s}(e^{-t/\tau_\mathrm{dm}}-1),\label{Eq:FitDemagn}
\end{equation}
yielding the characteristic time $\tau_\mathrm{dm}$=500$\pm$5\,ps.
The value $\theta_\mathrm{dm}/\theta_s$ characterizes the magnitude of this process and amounts to 0.2\%.
Noting that the magnetization is directed along the sample normal, we conclude from the experimental data that the action of the pump pulses induces slow exponential change of the magnetization magnitude.

The long time scale of 500\,ps and small magnitude of 0.2\,\% of the observed change of the magnetization suggest that pump pulses trigger the demagnetization in the sample.
Such a process in magnetically-ordered dielectrics was reported in Ref.\,\onlinecite{Kimel-PRL2002} and studied for the case of ferro- and ferrimagnetic,\cite{Hansteen-PRB2006,Ogasawara-PRL2005} weak ferromagnetic,\cite{Kimel-PRL2002} and antiferromagnetic\cite{Bossini-PRB2014} materials.
This process can be understood as follows.
Optical absorption of a laser pulse leads to excitation of electrons to the $3d$ sublevels of the Fe$^{3+}$ ions which decays at the femtosecond timescale leading to nonequilibrium phonon excitation and subsequent increase of the lattice temperature.\cite{Scott-PRB1974}
Relatively weak phonon-magnon interaction mediates energy transfer from lattice to incoherent magnons increasing thus the effective spin temperature. This appears as a decrease of magnetization magnitude.

We would like to note, that the slow change of the magnetization value can, in general, lead to the deviation of the magnetization from its equilibrium orientation resulting in a change of $M_z$.
Indeed, the orientation of the magnetization is defined by the balance between the magnetic anisotropy energy $w_\mathrm{a}$, the shape anisotropy energy $-4\pi M_z^2$, and the Zeeman energy $-\mathbf{M}\cdot\mathbf{H}_\mathrm{ext}$.
While the anisotropy energy depends on the normalized components $m_k$ of the magnetization but not on their absolute values (see Eq.\,(\ref{Eq:anisotropy})), the two other energies would decrease as the absolute magnitude of the magnetization decreases.
As a result, the relative contribution of the magnetic anisotropy $w_a$ grows.
If the easy-axis anisotropy energy is comparable to the Zeeman energy the demagnetization would lead to the increase of $M_z$ competing with decrease of $M_S$.
The described process is not expected to give any noticeable contribution in the film under consideration due to the dominance of the easy-axis anisotropy.

\end{document}